\documentstyle[aps,prl,twocolumn,epsf,epsfig]{revtex}
\begin{document}
\draft

\wideabs{
\title{Sensitivity to perturbations in a quantum chaotic billiard}
\author{Diego A. Wisniacki$^{1}$, Eduardo G. Vergini$^{1}$,
Horacio M. Pastawski$^{2}$ and Fernando M. Cucchietti$^{2}$}
\address{$^{1}$ Departamento de F\'{\i}sica, Comisi\'{o}n Nacional
de Energ\'{\i}a At\'{o}mica, 
Av. Libertador 8250, 1429 Buenos Aires, Argentina}
\address{$^{2}$ Facultad de Matem\'{a}tica, Astronom\'{\i}a y F\'{\i}sica,
Universidad 
Nacional de C\'{o}rdoba, 5000 C\'{o}rdoba, Argentina}
\date{\today}
\maketitle

\begin{abstract}
The Loschmidt echo (LE) measures the ability of a system to return to the
initial state after a forward quantum evolution followed by a backward
perturbed one.
It has been conjectured that the echo of a classically chaotic
system decays exponentially, with a decay rate given by the
minimum between the width $\Gamma$ of the local density of states and the
Lyapunov exponent.
As the perturbation strength is increased one obtains a cross-over between
both regimes.
These predictions are based on situations where the Fermi Golden Rule (FGR) is valid.
By considering a  paradigmatic fully chaotic system,
the Bunimovich stadium billiard, with a perturbation in a regime
for which the FGR  manifestly does not work, we find
a cross over from $\Gamma$ to Lyapunov  decay. We find that,
challenging the analytic interpretation,
these conjetures are valid even beyond the expected range.
\end{abstract}
} 

\narrowtext

Hypersensitivity to initial conditions is the key ingredient of classical
chaos. In quantum mechanics, its absence led to the study of other features
that could be associated with the chaos of the corresponding classical
system. Celebrated examples are the Gutzwiller trace formula for the quantum
spectral density, the description of the spectral fluctuations by the random
matrix theory and the relation of spectral correlations to 
transport \cite{cit-qchaos,cit-Szafer}.

In an alternative point of view, Peres \cite{cit-Peres} suggested that
quantum dynamics should distinguish regular and irregular classical dynamics
if the time evolution of an initial state for slightly different
Hamiltonians are compared. That is, the sensitivity of a quantum system
should be searched not by changing the initial conditions but rather by
perturbing the Hamiltonian. The natural quantity for this investigation is
the ability of the system to return to the initial state $\left| \phi
\right\rangle \,$ after being evolved with a Hamiltonian ${\cal H}_{0}$ for
a period $t$\ followed by an identical period of unitary evolution with $-%
{\cal H}_{1}=-( {\cal H}_{0}+\Sigma ).$ This defines the quantum Loschmidt
echo (LE)
\begin{equation}
M(t)=|\left\langle \phi \right| \exp [{\rm i}{\cal H}_{1} t/\hbar ]\exp [-
{\rm i}{\cal H}_{0}t/\hbar ]\left| \phi \right\rangle |^{2}.
\label{Eq-overlap}
\end{equation}
The perturbation $\Sigma $ can represent the uncontrolled degrees of freedom
of an environment. As in classical chaos, the LE is related to a `distance'
between a perturbed and an unperturbed evolution of the same initial state.

In recent years new hints were available due to the advances in Nuclear
Magnetic Resonance. The LE was measured in  a {\it %
many-body} system of interacting spins \cite{cit-polecho} in a range where
it is known to have spectral signatures of chaos.
A striking finding was that, when interaction with the
environment and residual interactions are very weak, the decay of $M(t)$
becomes independent of the perturbation strength. In this situation, it
depends on the dynamical scales of the systems, i.e. on ${\cal H}_{0}$.
While the complexity of the experimental system did not allow for a
derivation of the characteristic time for these specific system, Jalabert
and Pastawski \cite{cit-Jal-Pas} studied the LE in a {\it one-body}
classically chaotic Hamiltonian with a perturbation represented by a long
range quenched disordered potential. They have showed analytically that $M(t)
$\ may decay exponentially with a rate given by the Lyapunov exponent of the
classical system. As condition, the perturbation must be quantically strong
to produce statistically unpredictable changes in the quantum phase but weak
enough to leave the underlying classical dynamics undisturbed.

More recently, Jacquod, Silvestrov and Beenakker \cite{cit-Jacquod}
predicted a cross-over from a perturbation dependent regime to the Lyapunov
one. However, this prediction is based on the strong assumption that the
perturbation lives in a FGR regime; i.e. the local density of states (LDOS)
is a Breit-Wigner distribution whose width $\Gamma $ varies quadratically
with the perturbation strength. In this situation, $M(t)$ for a wave packet
and the survival probability of an unperturbed eigenstate have a decay rate
given by $\Gamma .$ Both observables would describe the same physics if the
correlation between states forming the wave packet could be neglected.

Our aim is to determine whether the perturbation independent
Lyapunov regime and the cross-over from a $\Gamma $ decay are possible in a
fully chaotic system with a clear semiclassical description where the
presence of the perturbation is not described by the FGR. This occurs when
there are strong correlations that could be related to classical structures
which prevents a description in terms of a random matrix theory. This
perturbation is then said to be {\it non generic} \cite{cit-doron1} and the
LDOS can be very different from the Lorentzian analyzed in Ref. \cite
{cit-Jacquod}. Our positive answer in such a case opens the question of a
semiclassical interpretation for the weak perturbation regime.

\begin{figure}[tbp]
\centering \leavevmode
\epsfxsize=4cm
\center{\epsfig{file=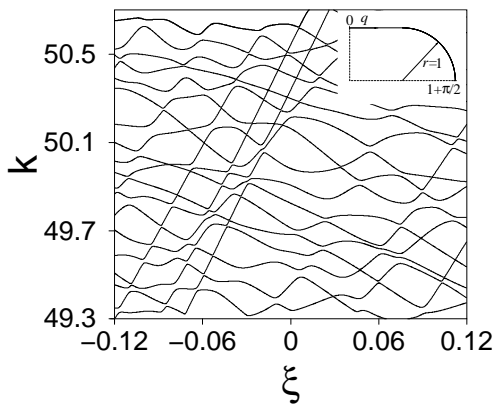, ,width=6.5cm,angle=0}}
\vspace{0.5cm}
\caption{Spectrum of the desymmetrized stadium billiard with mixed boundary
conditions controled by the parameter $\protect\xi$ [Eq. (3)]. The wave
numbers $k_{\protect\mu}(\protect\xi)$ run between 49.3 and 50.7. Inset:
Schematic figure of the system. In solid line we show the boundary of the
stadium billiard where the mixed boundary conditions are applied [Eq. (2)].
The coordinate $q$ on the boundary is also shown. Dashed lines correspond to
the symmetries axis with Dirichlet boundary conditions. }
\label{fig1}
\end{figure}

We consider the paradigmatic disymmetrized Bunimovich stadium
billiard \cite{cit-bunimovich}. It consists of a free particle inside a
2-dimensional planar region whose boundary ${\cal C}$ is shown in Fig. ~\ref
{fig1}. The radius $r$ is taken equal to unity and the enclosed area is $%
1+\pi /4$. This system not only has a great experimental relevance \cite
{cit-marcus-dephas-dot,cit-marcus-def-dot}, but also it is fully chaotic one
by oposition to the system considered in Ref.\cite{cit-Jacquod}. Besides, it
can rule out the diffusive effect of disorder suspected to affect the
behavior of $M(t)$ in a Lorentz gas\cite{cit-Lorentz}. The classical
dynamics is completely defined once the boundary is given. On the other
hand, to address the quantum mechanics, it is necessary to solve the
Helmholtz equation, $\nabla ^{2}\phi _{\mu }=k^{2}\phi _{\mu }$ with
appropriate boundary conditions. $k_{\mu }$ is the wave number and by
setting $\hbar =2m=1$, $k_{\mu }^{2}$ results the energy. The most commonly
used boundary conditions are the Dirichlet (hard walls) and the Neumann
(acoustics) conditions. However, we are interested in the possibility of
perturbing the quantum system without breaking the orthogonal symmetry and
leaving the classical motion undisturbed \cite{cit-Szafer}. This is possible
using more generalized boundary conditions:
\begin{equation}
\phi (q)\;+\;\xi \;g(q)\frac{\partial \phi }{\partial {\bf n}}(q)=0,
\label{mix}
\end{equation}
where $q$ is a coordinate along the boundary of the billiard (see Fig. ~\ref
{fig1}), and ${\bf n}$ is the unit vector normal to the boundary. $g(q)$ is
a real function and $\xi $ the parameter controlling the strength of the
perturbation. Dirichlet boundary conditions are recovered when $\xi =0$
while Neumann conditions are satisfied in the limit 
$\xi \rightarrow \infty $. 
The eigenfunctions and eigenenergies for the case $\xi =0$ are readily
obtained by using the scaling method\cite{cit-Vergini-Saraceno}.

In order to compute the LE in this system, a relation between the
eigenvalues and eigenfunctions for different values of the parameter $\xi $
is needed. Based on a recently developed Hamiltonian expansion for deformed
billiards \cite{cit-Wisniacki1}, it is easy to show that the eigenvalues and
eigenfunctions for different values of the parameter $\xi $ can be obtained
from the Hamiltonian ${\cal H}_{0}+\Sigma (\xi )$ which is expressed in the
basis of eigenstates at $\xi =0$ (from now on we will call $\phi _{\mu }$ to
these eigenstates),
\begin{equation}
\Sigma _{\mu \nu }^{{}}=\xi \times {\rm \Phi }_{\mu \nu }\;\oint_{{\cal C}%
}g(q)\;\frac{\partial \phi _{\mu }}{\partial {\bf n}}\frac{\partial \phi
_{\nu }}{\partial {\bf n}}{\rm d}q.  \label{Eq-Sigma-boundary}
\end{equation}
The function $g(q)$ measures the strength of the change in the boundary
condition along the contour. Within a perturbation theory it
would represent the direction and strength of a distortion of the stadium
\cite{cit-Wisniacki1}. Here we use
\[
g(q)=\left\{
\begin{array}{ccc}
\alpha &  & 0\leq q\leq 1, \\
(1+\alpha )\sin (q-1)+\alpha &  & 1<q\leq 1+\pi /2
\end{array}
\right.
\]
with $\alpha =-1/(2+\pi /2)$ that could be assimilated to a dilation along
the horizontal axis and a contraction along the perpendicular one. \ Notice
that the integral above could be viewed as an inner product among the wave
functions $\frac{\partial \phi _{\mu }}{\partial {\bf n}}$ defined over $%
{\cal C}$. This relation defines an effective Hilbert space in a window $%
\Delta k \approx$ Perimeter/Area \cite{cit-Wisniacki1}. The cut-off function
${\rm \Phi }_{\mu \nu }=\exp \left[ -2\;(k_{\mu }^{2}-k_{\nu
}^{2})^{2}/(k_{0} \Delta k)^{2}\right] $ restricts the effect of the
perturbation to states in this energy shell of width $B\simeq k_{0} \Delta k$%
. It allows to deal with a basis of finite dimension with wave numbers
around the mean value $k_{0}$ and restricting to a particular region $\Delta
k$ \ of interest.

Figure ~\ref{fig1} shows \ the dependence of the energy levels on the
perturbation. They exhibit many avoided crossings as $\xi $ is
varied. While the energy levels show the typical behavior of a general
system without constants of motion, we also recognize that some small
avoided crossings are situated along parallel tilted lines. These energies
correspond to the well known ``bouncing ball'' states which are highly
localized in momentum. The selected perturbation does not modify
substantially those states.

Since the LE is a classically motivated quantity,  a Gaussian wave
packet (with a mean value of momentum $k_{0}$ and velocity $v_{0}$) is
a proper semiclassical selection for an initial condition. By evaluating its
evolution in a system without perturbation ($\xi =0$) and other with
perturbation strength $\xi $, we compute the LE (Eq. 1) as a function of
time. At this point one must recognize that the choice of a semiclassical
initial condition is very relevant in order to observe the 'Lyapunov' regime
\cite{cit-Wisniacki2,cit-Jacquod}.

\begin{figure}[tbp]
\centering \leavevmode
\epsfxsize=4cm
\center{\epsfig{file=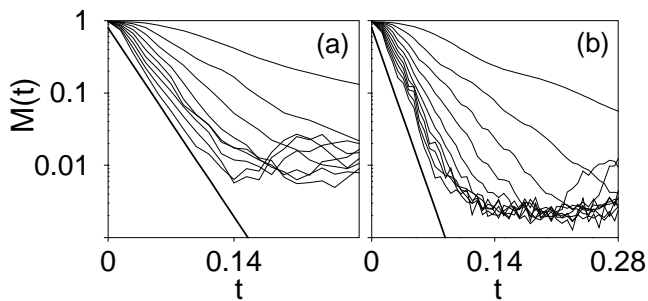, ,width=9.cm,angle=0}}
\vspace{0.5cm}
\caption{$M(t)$ for the desymmetrized stadium billiard perturbed by a change
in the boundary conditions. The calculations is shown in two different
energy regions. (a) Corresponds to the region around $k_{0}=50$. The value
of $\protect\xi $ is, from the top curve to bottom: 0.019, 0.038, 0.057,
0.075, 0.094, 0.11, 0.13, 0.15 and 0.17. (b) Corresponds to the region
around $k_{0}=100$. The value of $\protect\xi $ is, from the top curve to
bottom: 0.0066, 0.0131, 0.020, 0.0262, 0.0327, 0.0393, 0.0458, 0.0524,
0.0589, 0.066, and 0.072. The thick lines corresponds to an exponential
decay with decay rate $\protect\tau_{\protect\phi}=1/\protect\lambda$. }
\label{fig2}
\end{figure}

While a global exponential decay of $M(t)$ can be clearly identified in
almost any individual initial condition, the fluctuations for a system with $%
k_{0}$ not too large can introduce error in the estimation of the rate.
Hence, we have taken an average over $30$ initial states. Fig. (\ref{fig2})
(a) and (b) show typical sets of curves of $M(t)$ for $k_{0}=50$ and $%
k_{0}=100$ respectively. It can be seen that after a transient, $M(t)$
decays exponentially, $\sim \exp [-t/\tau _{\phi }]$. For $\xi >\xi
_{c}\simeq 4.5/k$ $\ $the decay rate $\tau _{\phi }$ becomes independent of
the perturbation and $1/\tau _{\phi }\approx \lambda $ with $\lambda $ the
Lyapunov exponent of the classical system \cite{cit-lyapunov} in accordance
with the conjecture. $\,\,$On the other hand, for large times $M(t)$
saturates to a finite value $M_{\infty }\approx 1/N$ with $N$ the effective
dimension of the Hilbert space \cite{cit-Peres}. $\,$

According to Ref. \cite{cit-Jal-Pas} the chaos controlled decay appears
provided that $\lambda >1/\widetilde{\tau }$ where $\widetilde{\ell }=v_{o}%
\widetilde{\tau }$ is the length over which the
perturbation changes the quantum phase (mean free path) which, for a plane
wave with wave number $k$ and velocity $v_{o}$. For a quenched disorder
perturbation is evaluated from the FGR \cite{cit-Jal-Pas}

\begin{equation}
\frac{1}{\widetilde{\tau }_{k}}=\frac{2\pi }{\hbar }\lim_{\eta \rightarrow
0^{+}}\sum_{k^{\prime }}\left| \Sigma _{k^{\prime }k}\right| _{{}}^{2}{%
\textstyle{\frac{1}{\pi }}}\frac{\eta /2}{(E_{k^{\prime }}-E_{k})^{2}+(\eta
/2)^{2}}.  \label{Eq_tau-FGR}
\end{equation}

Ref. \cite{cit-Jacquod} realized that in the opposite regime of $\lambda <1/%
\widetilde{\tau },$\thinspace\ \ the LE \ of an eigenstates $\phi _{\mu }$
of ${\cal H}_{0}$ is just a survival probability an must decay exponentially
under the action of the perturbation,
\begin{equation}
|\left\langle \phi _{\mu }\right| \exp [{\rm i}\left( {\cal H}_{0}{\cal +}%
\Sigma \right) t/\hbar ]\left| \phi _{\mu }\right\rangle |^{2}\sim \exp [-t/%
\widetilde{\tau }_{\mu }],  \label{Eq-Eigen-FGR}
\end{equation}
given by Eq. \ref{Eq_tau-FGR} for$\,\,\,\hbar /B<t<\hbar /\Delta $ ($\Delta $
the mean level spacing) \cite{cit-ferCOOKafa}. The appearance of this FGR
behavior requires that a typical matrix element $U\simeq \left\langle \left|
\Sigma _{\mu \nu }\left( \xi \right) \right| \right\rangle _{{\rm typ.}}$of
the perturbation to be $U>\Delta .$ The Fourier transform of Eq. \ref
{Eq-Eigen-FGR} is the LDOS which, although being discrete, would present a
Lorentzian envelope \cite{cit-Jac-FGR} of width $\Gamma =1/\widetilde{\tau }$%
. In Ref. \cite{cit-Jacquod} it is conjectured that this decay can determine
the LE decay with more general initial states. This is the regime controlled
by the non-diagonal terms in the semiclassical expansion \cite{cit-Jal-Pas}.
Once the non-diagonal terms have decayed, one expects the chaos controlled
decay of the diagonal ones will survive. This gives a cross-over criterion
for the decay rate of the LE of $1/\tau _{\phi }=\min [\Gamma ,\lambda ]$ as
the perturbative parameter $\xi $ changes.

\begin{figure}[tbp]
\epsfxsize=4cm
\center{\epsfig{file=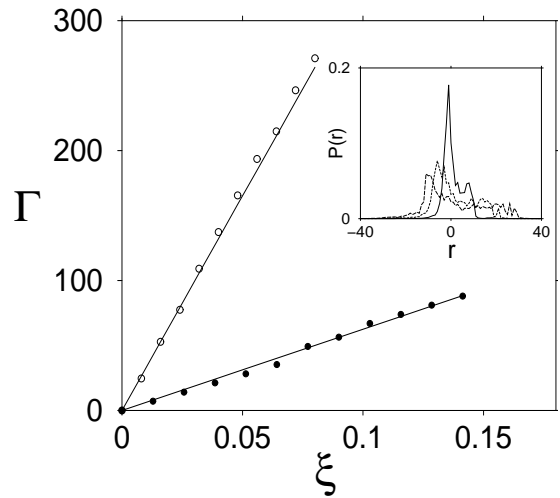, ,width=7.5cm,angle=0}}
\vspace{0.5cm}
\caption{Width $\Gamma$ of the local density of states as a function of the
perturbation strenght $\protect\xi$ for $k_{0}=50$ (filled circles) and $%
k_{0}=100$ (circles). The solid lines are the best linear fit. Inset: Local
density of states $P(r)$ for diferent perturbations in $k_{0}=50$ (r is
measured in mean level spacing units).}
\label{fig3}
\end{figure}

\begin{figure}[tbp]
\centering \leavevmode
\epsfxsize=4cm
\center{\epsfig{file=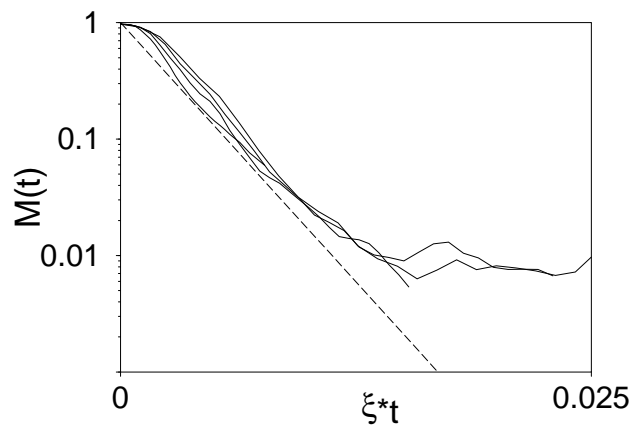, ,width=5.5cm,angle=-90}}
\vspace{0.5cm}
\caption{$M(t)$ as a function of the rescaled time $\protect\xi t$ for $%
k_{0}=50$ and $\protect\xi =$ 0.019, 0.038, 0.057 and 0.075. The dotted line
gives the decay $M(t)=\exp (-\Gamma t/2)$.}
\label{fig4}
\end{figure}

The LDOS is shown in Fig. ~\ref{fig3} for three different
perturbation strengths. In contrast to the case of Ref. \cite{cit-Jacquod}
our distribution is not Lorentzian. This is related to
the fact that the used perturbation (the function $g(q)$) does not connect
all different regions of phase space; for instance, the bouncing ball states
are practically undisturbed by $\Sigma $ determining the non-generic nature
of the perturbation. In particular, we have evaluated the width $\Gamma ,$
showing the spreading of the unperturbed eigenstates when expressed in terms
of the new ones. The results show a {\bf linear} dependence of $\Gamma $ on $%
\xi \,$ shown in Fig. 3; that is, we obtain $\Gamma \simeq 0.36\xi k^{2}$. \
Moreover, taking into account that $\lambda \simeq 0.86k$, the critical
value $\xi _{c}$ for the crossover from the $\Gamma $ regime to the Lyapunov
one is expected at $\xi _{c}=2.4/k$ (remember that from Fig. ~\ref{fig2} it
results $\xi _{c}\approx 4.5/k$). Then, for our system, the criterium works
with a $\Gamma $ given by the {\it half} width of the LDOS. This is shown in
Fig. ~\ref{fig4} where for perturbation strengths $\xi <4.5k^{-1}$, the LE%
{\bf .} decays as $M(t)=\exp \left[ -t/\tau _{\phi }\right] \,\,\,1/\tau
_{\phi }=\Gamma /2$ for $\lambda >\Gamma /2$.

These results contrast with the FGR dependence of $1/\tau _{\phi }\propto
\xi ^{2}$\ observed  for weak perturbations. These are the Lorentz gas with
a perturbed effective mass \cite{cit-Lorentz}, \ the kicked top perturbed by
a perpendicular delayed kick\cite{cit-Jacquod} and general chaotic system
perturbed by a quenched disorder\cite{cit-Jal-Pas} where random matrix
theory describes\cite{cit-cook-caio} the $\Gamma $ decay. In this context,
the linear dependence of $1/\widetilde{\tau }$ on $\xi $ may be considered
as a further indicative that the physics of the LE decay can be
very different from that described by Eq. \ref{Eq-Eigen-FGR} and that the
result of Ref. \cite{cit-Jacquod} has more general validity than expected.
In the non-perturbative regime, before the Lyapunov exponent takes over,{\it the LE} decays exponentially with a rate {\it given by the
perturbation dependent width of the LDOS.} Another important feature
 is that $\xi _{c}\simeq 4.5/k\,\rightarrow 0$ when $%
k\rightarrow \infty .$ This confirms that in the classical limit Eq. (\ref
{Eq-overlap}) would decay with the Lyapunov regime regardless of the
magnitude of $\Sigma$, recovering the chaotic hypersensitivity to
perturbations.

In summary, by studying one of the most important models in quantum chaos, a
fully chaotic billiard system, we have shown that, for a wide range of
parameters, the Loschmidt echo decays exponentially with rate given by the
Lyapunov exponent of the classical system. Moreover, we have discussed the
onset of this Lyapunov regime requires that $\lambda >\Gamma /2.$ In the
opposite situation, the presence of an exponential controlled by $\Gamma $
even in absence of a generic perturbation described by the FGR, demands
further studies to fully interpret the detailed mechanism controlling this
regime. We finally remark that the $M(t)$ would behave much differently for
intrisically quantum initial conditions. For an eigenstate of ${\cal H}$ one
finds a decay described by a FGR and it does not show a crossover into the
Lyapunov decay \cite{cit-Wisniacki2}.  In
the other quantum extreme, an initial state generated from the long time
evolution of a semiclassical wave packet\cite
{cit-polish}, we find a perturbation dependent Gaussian decay \cite
{cit-cook-gaussian}. These issues have begun to receive much attention \cite
{cit-preprints} due to its strong connection with quantum computing
stability, decoherence in waves, and quantum-classical transition.
Furthermore, the dephasing time observed in transport experiments in
mesoscopic devices shows a perturbation independent rate \cite
{cit-marcus-dephas-dot}. So far, there is no consensus about the physical
phenomenon causing it. Since the time scale $\tau _{\phi }$ measured by the
LE is a decoherence time and our methodology can obviously be adapted to
treat the transport problem \cite{cit-Szafer}, our results open a rich field
for exploration: the connection of both time scales.

We thank D. Cohen, R. Jalabert and M. Saraceno for
very useful discussions and the support from SeCyT-UNC, CONICET, ANPCyT,
ECOS-SeTCIP and Antorchas-Vitae. DAW received support from CONICET
(Argentina) and AECI (Spain).


\begin{references}
\bibitem{cit-qchaos}  H.--J. St\"{o}ckmann, {\it Quantum Chaos: An
Introduction} (Cambridge U.\ Press, Cambridge, 1999).

\bibitem{cit-Szafer}  A. Szafer and B. Altshuler, Phys. Rev. Lett. {\bf 70,}
587 (1993).

\bibitem{cit-Peres}  A. Peres, Phys. Rev. A {\bf 304,} 1610 (1984).

\bibitem{cit-polecho}  H. M. Pastawski, G. Usaj and P. R. Levstein, Chem.
Phys. Lett. {\bf 261,} 329 (1996);  H.\ M. Pastawski, P. R.
Levstein, G. Usaj, J. Raya and J. Hirschinger, Physica A {\bf 283,} 166
(2000).


\bibitem{cit-Jal-Pas}  R. Jalabert and H. M. Pastawski , Phys. Rev. Lett.
{\bf 86,} 2490 (2001).

\bibitem{cit-Jacquod}  Ph. Jacquod, P. G. Silvestrov, C. W. J. Beenakker,
Phys. Rev. E {\bf 64,} 055203 (2001)

\bibitem{cit-doron1}  D. Cohen, A. Barnett and E.J. Heller, Phys. Rev. E
{\bf 63}, 46207 (2001).

\bibitem{cit-bunimovich}  L. A. Bunimovich, Funct. Anal. Appl. {\bf 8,} 254
(1974).

\bibitem{cit-marcus-dephas-dot}  A. G Huibers et al. Phys. Rev. Lett. {\bf %
83,} 5090 (1999)

\bibitem{cit-marcus-def-dot}  M. Switkes, C. M. Marcus, K. Campman and A. C.
Gossard, Science {\bf 283}, 1905 (1999).

\bibitem{cit-Lorentz}  F. M. Cucchietti, H. M. Pastawski and D. A.
Wisniacki, Phys. Rev. E \ in press, cond-mat/0102135.

\bibitem{cit-Vergini-Saraceno}  E. \ Vergini and M. Saraceno, Phys. Rev. E
{\bf 52}, 2204 (1995).

\bibitem{cit-Wisniacki1}  D. A. Wisniacki and E. Vergini, Phys. Rev. E {\bf %
59}, 6579 (1999).

\bibitem{cit-Wisniacki2}  D. A. Wisniacki and D. Cohen, quant-ph/0111125.

\bibitem{cit-doron2}  D. Cohen and E.J. Heller, Phys. Rev. Lett {\bf 84},
2841 (2000).

\bibitem{cit-lyapunov}  Ch. Dellago and H. A. Posch, Phys. Rev. E {\bf 53,}
2401 (1995).

\bibitem{cit-ferCOOKafa}  F. M. Cucchietti, H. M. Pastawski, E. Medina and G
Usaj, Anales AFA {\bf 10, }224 (1998).

\bibitem{cit-Jac-FGR}  Ph. Jacquod \ and D. L. Shepelyansky, Phys. Rev.
Lett. {\bf 75}, 3501 (1995)

\bibitem{cit-cook-caio}  F. M.Cucchietti et al., nlin.CD/0112015.

\bibitem{cit-polish}  Z. P. Karkuszewski, C. Jarzynski, W. H. Zurek.
quant-ph/0111002

\bibitem{cit-cook-gaussian}  F. M. Cucchietti, et al. (unpublished)

\bibitem{cit-preprints}  P. G. Silvestrov, H. Sch\"{o}merus, and C. W. J.
Beenakker, Phys. Rev. Lett. {\bf 86} 5192 (2001); T. Prosen and M. Znidaric.
nlin.CD/0111014; N. R. Cerruti and S. Tomsovic. Phys. Rev. Lett. {\bf 88} 
054103 (2002); T. Gorin and T. H. Seligman, quant-ph/0112030.
\end{references}
\end{document}